\documentclass[prd,aps,twocolumn,nofootinbib,superscriptaddress,eqsecnum,floatfix,preprintnumbers,amsmath,amssymb,
nofootinbib,longbibliography]{revtex4-1}

\usepackage{amssymb,amsmath}
\usepackage{epsfig}
\usepackage[dvipsnames]{xcolor}
\usepackage[utf8]{inputenc}
\usepackage{stmaryrd}
\usepackage{mathrsfs}
\usepackage{mathalfa}
\usepackage{accents}
\usepackage[normalem]{ulem}
\usepackage{enumitem}

\usepackage{graphicx}

\newcommand{\be}{\begin{equation}}
\newcommand{\ee}{\end{equation}}
\newcommand{\ba}{\begin{eqnarray}}
\newcommand{\ea}{\end{eqnarray}}
\newcommand{\non}{\nonumber}
\newcommand{\n}[1]{\label{#1}}
\newcommand{\eq}[1]{(\ref{#1})}
\newcommand{\hh}{\, ,\hspace{0.25cm}}
\newcommand{\hhh}{\, ,\hspace{0.5cm}}
\newcommand{\BM}[1]{{\mbox{\boldmath $#1$}}}
\newcommand{\ind}[1]{\mbox{\tiny #1}}

\newcommand{\bh}[1]{\bar{h}}

\begin{document}

\title{Gravitational spinoptics in a curved space-time}

\author{Valeri P. Frolov}%
\email[]{vfrolov@ualberta.ca}
\affiliation{Theoretical Physics Institute, Department of Physics,
University of Alberta,\\
Edmonton, Alberta, T6G 2E1, Canada
}
\author{Andrey A. Shoom}
\email[]{ShAAndrey@kantiana.ru}
\affiliation{Immanuel Kant Baltic Federal University, 236016 Kaliningrad, Russia}

%\today
\begin{abstract}
In this paper we discuss propagation of the  weak high-frequency  gravitational waves in a curved spacetime background. We develop a so-called spinoptics approximation which takes into account interaction of the spin of the field with the curvature of the background metric. This is achieved by modifying the standard geometric optics approximation by including the helicity sensitive terms of the order $1/\omega$ in the eikonal equation. The novelty of the approach developed in this paper is that instead of study of the high-frequency expansion of the equations for the gravitational field perturbations we construct the effective action for the gravitational spinoptics. The gravitational spinoptics equations derived by variation of the  effective action correctly reproduce the earlier obtained results. However, the proposed effective action approach is technically more simple and transparent.
It allows one to reduce the study of the high-frequency gravitational waves to study classical dynamics of massless particles with internal discrete degree of freedom (helicity). The formalism is covariant and it can be applied for arbitrary vacuum space-time background.

\medskip
\hfill {\scriptsize Alberta Thy 5-24}
\end{abstract}

\maketitle

\section{Introduction}

After the first discovery of the gravitational waves by LIGO in 2015 \cite{PhysRevLett.116.061102}, the gravitational wave channel  provides us now with a main tool for observation and study of the coalescence in black hole-black hole and black hole-neutron star  binary systems. In the General Relativity the gravitational waves are described by the metric perturbations propagating with the speed of light. Equations describing the gravitational waves are more complicated than the Maxwell equations, however these waves have quite many common features with the electromagnetic waves. In particular, there exists two independent solutions for a monochromatic plane gravitational wave corresponding to their polarization states. Similarly to the electromagnetic waves, one can choose right- and left-circular polarized states as a basis in this two dimensional space. In both electromagnetic and gravitational cases such solutions correspond to states with a fixed helicity, equal $\pm 1$ for photons and $\pm 2$ for gravitons. The solutions for a linear and elliptic polarization states can be obtained as a linear superposition of the circular polarized solutions.

A powerful method of study high-frequency electromagnetic and gravitational waves is a well-known geometric optics approximation. The basic idea of this approach is that when the wavelength of the radiation is much less than other characteristic length parameters (such as the curvature of the wavefront and the spacetime curvature), one can approximate a solution by assuming that locally it is similar to the monochromatic plane wave. Formally this is achieved by expanding the wave solutions in the inverse powers of the frequency and keeping
a few lowest order terms. In the application of the geometric optics approximation  for the electromagnetic and gravitational linearly polarized waves propagating in a curved spacetime 
the results are the following. In both cases, solutions of the eikonal equations are generated by null geodesics, and the vector or tensor associated with the linear polarization is parallel propagated along these null geodesics. The number of photons or gravitons inside "tubes" formed by 2D set of null geodesics passing through a small 2D sphere is conserved. These are the main results of the leading order of the $1/\omega$ expansion. Details and corresponding references can be found in the book \cite{MTW}.

In the next, first order in $1/\omega$ expansion one can demonstrate that the plane of the polarization of the beam of the photons or gravitons associated with the geometric-optics solution slowly rotates with respect to an external observer. This a so called effect of Faraday rotation. Faraday effect for the electromagnetic and gravitational waves in a curved spacetime background was discussed in a number of publications (see e.g. \cite{Skrotskii,Plebanski,God,GF2,GF3,GF4,GF5,Piran:1985,GF6,CariniRuffini,Perlick_1993,GF7,GF8,GF9,Halilsoy:2006ev,GF10,Frolov:2011mh,Ghosh,Dolan:2018nzc,Dolan:2018ydp,Hou:2019wdg,Li:2022izh,Shoom:2022oer} and references therein).

A described Faraday rotation effect is a result of the interaction of the spin of photons and gravitons with the curvature of the spacetime. One can expect that there should exists a counterpart of this effect. Namely, curvature acting on the spin should modify a motion of a spin carrying objects, photons or gravitons. This modification is of the order of $1/\omega$, and hence the eikonal equation of the standard geometric optics is not sensitive to it. In order to study this effect one should modify the geometric optics. The idea of this modification, often called spinoptics approach, is the following. Instead of working with linear polarized solutions, one focuses on the propagation of the circular polarized solutions with a fixed helicity. One can observe that in the first order of the $1/\omega$ expansion there exist special helicity dependent terms. Namely these terms are responsible for the change of the trajectory, and they can play an important role at large distance. To take into account this long-distance effect one should "enhance" the helicity-dependent terms and include them in a modified eikonal equation. As a result, the set of rays associated with the high-frequency wave remains null, but loses the property to be geodesic.  This approach was developed for the electromagnetic waves in the publications \cite{Frolov:2011mh,Frolov:2012zn,Yoo:2012vv,Oancea:2019pgm,Oancea:2020khc,Frolov:2020uhn,Dahal:2022gop}. Spinoptic approach for the gravitational waves in a curved spacetime was discussed in
\cite{Yamamoto:2017gla,Andersson:2020gsj,Dahal:2021qel,Kubota:2023dlz}.
Let us emphasize that in the papers \cite{Oancea:2019pgm,Oancea:2020khc,Frolov:2020uhn}
the spinoptics equations were derived in an arbitrary vacuum spacetimes, without additional assumptions about its structure. A covariant approach to the gravitational Faraday effect and dual to it the spin-Hall effect of light in arbitrary vacuum spacetimes was presented in \cite{Shoom:2020zhr}, without an explicit observer field specified, and a local description of these effects was given in \cite{Shoom:2024zep}.

The "standard" derivation of the electromagnetic and gravitational spinoptics equations is based on a detailed analysis of the  high-frequency expansions of the corresponding wave equations. Recently, a new method of derivation of the electromagnetic spinoptics equations was developed in \cite{Frolov:2024ebe}. This method is based on the old Stachel and Plebanskii's idea to use an effective action approach for the high-frequency approximation for the particles with spin \cite{Stachel:1977cm}. In paper \cite{Frolov:2024ebe} an effective action for the electromagnetic spinoptics  is derived and used for study of the propagation of the  high-frequency electromagnetic waves in a curved spacetime. The purpose of this paper is to demonstrate that a similar method
can be used for the high frequency gravitational waves.  Namely, we derive the corresponding effective action in the gravitational spinoptics approximation, obtain gravitational spinoptics equations and study their properties. 

The paper is organized as follows. In Section~II we briefly discuss plane monochromatic gravitational waves. Section III describes the  action for weak gravitational waves propagating in an arbitrary vacuum spacetime.  In Sec.~IV we derive an effective action for the gravitational spinoptics formalism.  In section~V we present a derivation of the gravitational spinoptics equations and discuss  their properties. Section~VI contains a discussion of the obtained results. Appendix contains a brief summary  of the derivation of the second order perturbation of the Einstein-Hilbert action.

We use a system of units in which $G=c=1$ and sign conventions adopted in the book \cite{MTW}.

\section{Monochromatic plane gravitational waves}

We begin with a brief review of  weak plane monochromatic  gravitational waves propagating in a flat (Minkowski) spacetime. Such a wave is described by a solution of the linearized vacuum Einstein equations and has the form
\be\n{0.1a}
h_{\mu\nu}=\Re\{A_{\mu\nu}e^{ik_{\alpha}x^{\alpha}}\}\,.
\ee
Here
\be\n{0.2}
k^{\mu}k_{\mu}=0
\ee
is a null vector in the direction of the wave propagation, $\Re\{\ldots\}$ stands for the real part of $\{\ldots\}$, and $A_{\mu\nu}$, satisfying relations
\be\n{0.3}
A_{\mu\nu}k^{\nu}=0\hh A^{\mu}_{\;\mu}=0\, ,
\ee
is a tensor characterizing the waves amplitude and polarization.

Consider an inertial frame and let
$U^{\mu}$ be a  unit timelike vector of the observer at rest in this frame. The frequency $\omega$ of the monochromatic wave as measured by the observer $U^{\mu}$ is
\be
\omega=-U^{\mu}k_{\mu}\, .
\ee
The vector $k^{\mu}$ can be written in the form $k^{\mu}=\omega(1,\vec{n})$, where $\vec{n}$ is a unit 3D vector in the direction of the wave propagation. The 4D spacelike unit vector  orthogonal to $U^{\mu}$ is
\be
N^{\mu}=\dfrac{1}{\omega}k^{\mu}-U^{\mu}\, .
\ee
Two vectors, $U^{\mu}$ and $k^{\mu}$ (or $N^{\mu}$), span a two-dimensional timelike plane $\Gamma$. Denote by $\Pi$ a two-dimensional spacelike plane, orthogonal to $\Gamma$. We call it a screen plane.
Using the gauge ambiguity one can put $A_{\mu\nu}U^{\nu}=0$.

The tensor of the amplitude of the gravitational wave, satisfying the above described conditions can be reduced to a tensor "living" on the screen plane $\Pi$. It characterizes the polarization states of the wave.
To describe these states it is convenient to introduce basic vectors on the screen plane.
We denote them by $e^{\mu}_{\,\,\, 1}$ and $e^{\mu}_{\,\,\, 2}$. These two mutually orthogonal units vectors  span the screen plane $\Pi$. We choose the 4D basis $(U^{\mu},N^{\mu},e^{\mu}_{\,\,\, 1},e^{\mu}_{\,\,\, 2}) $ to be right-handed, that is satisfying the relation
\be
e_{\mu\nu\lambda\rho}U^{\mu}N^{\nu}e^{\lambda}_{\,\,\, 1}e^{\rho}_{\,\,\, 2}=1\, .
\ee
We also define the complex  null vectors
\be\n{0.4}
m^{\mu}=\frac{1}{\sqrt{2}}(e^{\mu}_{\,\,\, 1}+ie^{\mu}_{\,\,\, 2})\hh \bar{m}^{\mu}=\frac{1}{\sqrt{2}}(e^{\mu}_{\,\,\, 1}-ie^{\mu}_{\,\,\, 2})\,.
\ee

For the plane monochromatic gravitational wave one can define its unit linear-polarisation modes as follows:
\be\n{0.5}
e^{\mu\nu}_{\,\,\, +}=e^{\mu}_{\,\,\, 1}e^{\nu}_{\,\,\, 1}-e^{\mu}_{\,\,\, 2}e^{\nu}_{\,\,\, 2}\hhh e^{\mu\nu}_{\,\,\, \times}=e^{\mu}_{\,\,\, 1}e^{\nu}_{\,\,\, 2}+e^{\mu}_{\,\,\, 2}e^{\nu}_{\,\,\, 1}\,.
\ee
These modes are orthogonal and inclined to each other at $\pi/4$ angle. The unit circular polarisation modes are defined accordingly,
\ba\n{0.6}
e^{\mu\nu}_{\,\,\,\ind{R}}&=&\frac{1}{\sqrt{2}}(e^{\mu\nu}_{\,\,\, +}+ie^{\mu\nu}_{\,\,\, \times})=m^{\mu}m^{\nu}\, ,\non\\
e^{\mu\nu}_{\,\,\,\ind{L}}&=&\frac{1}{\sqrt{2}}(e^{\mu\nu}_{\,\,\, +}-ie^{\mu\nu}_{\,\,\,\times})=\bar{m}^{\mu}\bar{m}^{\nu}\,,
\ea
where the subscripts $\mbox{R}$ and $\mbox{L}$ stand for right-handed and left-handed gravitational polarisation modes. These polarisation modes correspond to $\sigma=+2$ and $\sigma=-2$ helicity states, respectively \cite{MTW}, which are conserved \cite{Barnett:2014era,Aghapour:2021bkb}.

Let us summarize: A solution of the linearized gravitational equations in the flat spacetime descibing a monochromatic plane wave with 4D wave vector  $\bf{k}$ and positive helicity  is
\be\n{0.1}
h_{\mu\nu}=\Re\{A m_{\mu}m_{\nu}e^{ik_{\alpha}x^{\alpha}}\}\,,
\ee
A similar expression with the change $m_{\mu}$ to $\bar{m}_{\mu}$ describes a wave with negative helicity\footnote{Let us note, that for the Maxwell theory the solutions describing monochromatic circularly polarized plane have similar form. Namely, $A_{\mu}=\Re \{a m_{\mu}e^{ik_{\alpha}x^{\alpha}}\}$ is an electromagnetic potential for the wave with helicity $\sigma=+1$, while $A_{\mu}=\Re \{a \bar{m}_{\mu}e^{ik_{\alpha}x^{\alpha}}\}$ describes a wave with the helicity $\sigma=-1$.}.

\section{Action for gravitational waves}

For derivation of the effective action in the spinoptics approximation one considers a metric  $\hat{g}_{\mu\nu}$ splitted into the background metric $g_{\mu\nu}$ and its perturbation $h_{\mu\nu}$ describing the gravitational waves
\be\n{1.1}
\hat{g}_{\mu\nu}=g_{\mu\nu}+h_{\mu\nu}\,.
\ee
One assumes that $h_{\mu\nu}$ is small. Then the quantities such as $\hat{g}^{\mu\nu}$, $\hat{g}=\mbox{det}(\hat{g}_{\mu\nu})$ and Ricci tensor $\hat{R}_{\mu\nu}$ can be written as a series expansion of powers of $h_{\mu\nu}$ and its derivatives. Keeping the terms up to the second order in $h_{\mu\nu}$ one has
\be
\begin{split}\n{sec}
&\hat{g}^{\mu\nu}={g}^{\mu\nu}+{g}^{(1)\mu\nu}+{g}^{(2)\mu\nu}\, ,\\
&\hat{g}=g+g^{(1)}+g^{(2)}\, ,\\
&\hat{R}_{\mu\nu}= R_{\mu\nu}+R^{(1)}_{\mu\nu}+R^{(2)}_{\mu\nu}
\,,
\end{split}
\ee
The terms with the superscript $(1)$ and $(2)$ contain one and two powers of the perturbation $h_{\mu\nu}$, respectively. Explicit form of the expressions in \eqref{sec} can be found in the section $\S 35$ of the book \cite{MTW} (see also Appendix).

Let us consider the Einstein-Hilbert action for the gravitational field,
\be\n{1.6}
\hat{I}=\frac{1}{16\pi}\int d^4x\sqrt{-\hat{g}}\,\hat{g}^{\mu\nu}\hat{R}_{\mu\nu}\, ,
\ee
and substitute expansions \eqref{sec} into it. This gives us the expansion
\be
\hat{I}=I_0+I_1+I_2\, .
\ee
This expansion of the gravity action in powers of the perturbation $h_{\mu\nu}$ up to its second order was obtained by MacCallum and Taub \cite{Maccallum:1972er}  (see also \cite{Stein:2010pn,OOH}).
For completeness and in order to fix notations we briefly reproduce the derivation of $\hat{I}$ in the appendix.

In what follows, we assume that the background metric $g_{\mu\nu}$ satisfies vacuum Einstein equations, $R_{\mu\nu}=0$, then the terms $I_0$ and $I_1$ vanish and $I_2$ greatly simplifies and takes the form
\be
\begin{split}\n{I2}
&I_2=\frac{1}{32\pi}\int d^4x\sqrt{-{g}}L\,,\\
&L=q^{\alpha\beta ;\gamma}q_{\beta\gamma ;\alpha}-\dfrac{1}{2}q^{\alpha\beta ;\gamma} q_{\alpha\beta ;\gamma}+\dfrac{1}{4}q^{;\alpha}q_{;\alpha}\, ,\\
&q=-h\equiv -g^{\mu\nu}h_{\mu\nu}\hh
q_{\mu\nu}=h_{\mu\nu}-\dfrac{1}{2}g_{\mu\nu}h\, .
\end{split}
\ee
In these expressions and later on all the operations with the indices are performed by using the background metric $g_{\mu\nu}$ and its inverse $g^{\mu\nu}$. The semicolon denotes a covariant derivative with respect to the background metric. The expression for  $I_2$ contains only first derivatives of the perturbation. This is achieved by
the integration by parts of the terms with second derivatives and omitting the corresponding boundary terms\footnote{
Let us mention that we use notation $q_{\mu\nu}$ for the quantities which are usually are denoted by $\bar{h}_{\mu\nu}$. The reason is that later we shall deal with the complexified version of the action and will denote by bar the operation of the complex conjugation. Written in terms of $h_{\mu\nu}$ the second order perturbation action \eqref{I2} coincides with \eqref{A10}. }.

The Euler-Lagrange equation obtained by varying action $I_2$ with respect to the perturbation $h_{\mu\nu}$
has the form\footnote{It is possible to check that this equation coincides with the equation $R^{(1)}_{\mu\nu}=0$ (see e.g. equation (35.58a) in \cite{MTW}). }
\be\n{qqqq}
q_{\alpha\beta\ ;\gamma}^{\ \  \  ;\gamma}-2q_{\gamma (\alpha  ;\beta)}^{\quad\quad ;\gamma}-\dfrac{1}{2}g_{\alpha\beta} q^{;\gamma}_{\ \ ;\gamma}=0\, .
\ee
This equation describes a propagation of the gravitational perturbations (waves) in a curved spacetime background with metric $g_{\mu\nu}$. Written in terms of $h_{\mu\nu}$ this equation has the form \eqref{A13}.

For the derivation of the effective action for the gravitational spinoptics we shall use a complexified version of the second order perturbed action $I_2$, Namely, following  \cite{Stachel:1977cm} we assume that the perturbation field $h_{\mu\nu}$ takes complex value and define the following action, which we denote by ${\cal W}$
\be
\begin{split}\n{WWW}
{\cal W}=\frac{1}{32\pi}\int d^4x\sqrt{-g}\Big( &
q^{\alpha\beta ;\gamma}\bar{q}_{\beta\gamma ;\alpha}
+\bar{q}^{\alpha\beta ;\gamma}q_{\beta\gamma ;\alpha}\\
&-
q^{\alpha\beta ;\gamma} \bar{q}_{\alpha\beta ;\gamma}+\dfrac{1}{2}q^{;\alpha}\bar{q}_{;\alpha}\Big)\, .
\end{split}
\ee
Here a bar denotes a complex conjugation.

\section{Effective action and its variations}

\subsection{Effective action}

We use the action \eqref{WWW} as our starting point for the derivation of the effective action for the gravitational spinoptics. For this purpose we consider the following ansatz for the complex perturbation field
\be\n{3.1}
h_{\mu\nu}=A M_{\mu}M_{\nu}e^{i\omega S}\,,
\ee
where $A$ is a real function describing the amplitude of the complex wave and $S$ is a real function called the eikonal. A constant frequency $\omega$ is assumed to be large and it will be used to track the order of the high-frequency expansion. In this expansion the $A$, $S$, and the polarization vector $M^{\mu}$ are assumed to be of the order of 1\footnote{For the high frequency complex wave $h_{\mu\nu}$ the complexified action \eqref{WWW} can be "derived" as follows. Consider $q_{\mu\nu}$ constructed for the sum $h_{\mu\nu}+\bar{h}_{\mu\nu}$  and substitute it in \eqref{I2}. Integrals containing the high frequency oscillating factors $e^{\pm 2i\omega S}$ can be neglected. What is left reproduces $\cal{W}$.
}.

Following \cite{Frolov:2024ebe}
we impose the following conditions
\be\begin{split}\n{MMM}
&M^{\mu}M_{\mu}=\bar{M}^{\mu}\bar{M}_{\mu}=0\, ,\\ &M^{\mu}\bar{M}_{\mu}=1\hh
M^{\mu}S_{;\mu}=0\, .
\end{split}
\ee
It is easy to see that \eqref{3.1}-\eqref{MMM} are similar to \eqref{0.1} with $M_{\mu}=m_{\mu}$ and $k_{\mu}=\omega S_{,\mu}$.

In the next step, we substitute the ansatz \eq{3.1} into the action \eq{WWW}. Let us note that for a chosen ansatz the trace of the complex perturbation $h_{\mu\nu}$ vanishes. This means that it is sufficient to keep the same form of the action ${\cal W}$ in which one simply puts $q_{\mu\nu}=h_{\mu\nu}$

Then expanding the action \eqref{WWW} in powers of $\omega$ and keeping  the first two leading order terms of this expansion, one gets
\be\n{3.10}
{\cal W}=-\frac{1}{16\pi}\int d^4x\sqrt{-g}\,\omega^2A^2\left(\frac{1}{2}(\nabla S)^2-\frac{2}{\omega}S_{;\mu}B^{\mu}\right)\,.
\ee
Here
\be\n{3.11}
B^{\alpha}=i\bar{M}_{\mu}M^{\mu;\alpha}
\ee
is a real vector. It is easy to see that this effective action \eqref{3.10} is similar to the action for the Maxwell field given in \cite{Frolov:2024ebe}. The only difference is that the second term in the integrand is twice larger.
The action \eq{3.10} depends on three variables: the amplitude $A$, vector $M^{\mu}$, and the eikonal function $S$. We define the effective action for the spinoptics approximation as ${\cal I}=-16\pi{\cal W}/\omega^2$, which differs from ${\cal W}$ by a constant factor.

In the derivation of ${\cal W}$ we used the relations \eq{MMM}. This implies that we have to supplement the action with the corresponding constraints,
\ba\n{3.12}
\Lambda &=&\dfrac{1}{2}\bar{\lambda}_1 M_{\mu}M^{\mu}+\dfrac{1}{2}\lambda_{1}\bar{M}_{\mu}\bar{M}^{\mu}+\lambda_{2} ({M}_{\mu}\bar{M}^{\mu}-1)\non\\
&+&\bar{\lambda}_{3} M^{\mu} S_{,\mu}+\lambda_{3}\bar{M}^{\mu}S_{,\mu}.
\ea
Here $\lambda_1$ and $\lambda_3$ are complex Lagrange multipliers, and $\lambda_2$ is a real one.
Including these constraints into the effective action ${\cal I}$ we obtain the following effective action for the gravitational spinoptics
\be\n{3.13}
{\cal I}=\int d^4x\sqrt{-g}\,\left[A^2\left(\frac{1}{2}(\nabla S)^2-\frac{2}{\omega}S_{;\mu}B^{\mu}\right)+\Lambda\right]\,.
\ee

\subsection{Variations of the effective action}

By taking variations of ${\cal I}$ with respect to $A$, $M^{\mu}$, and $S$ one obtains the Euler-Lagrange equations
describing the dynamics of our system.
\begin{itemize}
\item The variation $\delta{\cal I}/\delta A$ together with the condition $A\ne0$ leads to the following equation:
\be\n{3.16}
H\equiv\frac{1}{2}(\nabla S)^2-\frac{2}{\omega}S_{;\mu}B^{\mu}=0\,.
\ee
\item The variation $\delta{\cal I}/\delta S$ leads to the continuity equation
\be\n{3.17}
J^{\mu}_{\;\;;\mu}=0\,
\ee
for the current
\be\n{3.18}
J^{\mu}=A^2\left(S^{;\mu}-\frac{2}{\omega}B^{\mu}\right)+\bar{\lambda}_{3}M^{\mu}+\lambda_{3}\bar{M}^{\mu}\,.
\ee
\item The variation $\delta{\cal I}/\delta \bar{M}^{\mu}$ leads to
\be\n{3.19}
-\frac{2i}{\omega}A^2S^{;\nu} M_{\mu;\nu}+\lambda_{1} \bar{M}_{\mu}+\lambda_{2} M_{\mu}+\lambda_{3}S_{;\mu}=0\,.
\ee
\end{itemize}
The variation of the effective action with respect to $M^{\mu}$ gives the equation complex conjugated to \eqref{3.19} and hence it does not contain new information.

The relations \eq{MMM} are invariant under rotation in a 2D plane spanned by the complex null vectors $M^{\mu}$ and $\bar{M}^{\mu}$,
\be\n{3.14}
M^{\mu}\to e^{i\psi} M^{\mu}\hh \bar{M}^{\mu}\to e^{-i\psi} \bar{M}^{\mu}\,.
\ee
Under this transformation the vector $B^{\mu}$ changes as follows: $B_{\mu}\to B_{\mu}-\psi_{,\mu}$.
Using this gauge freedom we put
\be\n{3.15}
B^{\mu}S_{;\mu}=0\,.
\ee

Contracting this equation with $M^{\mu}$ and using \eq{MMM} leads to $\lambda_1=0$ and contracting it with $\bar{M}^{\mu}$ and using \eq{MMM}, \eq{3.11}, and \eq{3.15} leads to $\lambda_2=0$. Thus, equation \eq{3.19} takes the following form:
\be\n{3.20}
\frac{2i}{\omega}A^2S^{;\nu} M_{\mu;\nu}=\lambda_{3}S_{;\mu}\,.
\ee

\section{Spinoptics formalism}

\subsection{Hamilton-Jacobi equation and congruence of null rays}

Following the prescription of \cite{Stachel:1977cm} we identify \eq{3.16} with the Hamilton-Jacobi equation for a particle with 4-momentum $p_{\mu}=S_{;\mu}$.  The corresponding Hamiltonian reads
\be\n{3.21}
H(x^{\mu},p_{\mu})=\frac{1}{2}g^{\mu\nu}p_{\mu}p_{\nu}-\frac{2}{\omega}p_{\mu}B^{\mu}\,.
\ee
This Hamiltonian determines a class of mechanical trajectories (rays) associated with high-frequency gravitational waves. The corresponding Hamilton equations are the following:
\ba\n{3.22}
\frac{dx^{\mu}}{ds}&=&\frac{\partial H}{\partial p_{\mu}}=p^{\mu}-\frac{2}{\omega}B^{\mu}\,,\\
\frac{d p_{\mu}}{ds}&=&-\frac{\partial H}{\partial x^{\mu}}=-\frac{1}{2}g^{\alpha\beta}_{\;\;\;\;,\mu}\,p_{\alpha}p_{\beta}+\frac{2}{\omega}p_{\nu}B^{\nu}_{\;,\mu}\,.\non
\ea
Here $s$ is a parameter along particle trajectories. With the aid of the first Hamilton equation, the second one can be written in the covariant form
\be\n{3.23}
\frac{Dp_{\mu}}{ds}=\frac{2}{\omega}p_{\nu}B^{\nu}_{\;\;;\mu}\,.
\ee
Here $D/ds$ stands for the covariant derivative with respect to the background metric.

Let us now rewrite the Hamilton equations \eq{3.22}, \eq{3.23} in terms of kinematic quantities. Using \eq{3.22} we denote by $l^{\mu}$ a tangent vector to the ray $x^{\mu}=x^{\mu}(s)$,
\be\n{3.24}
l^{\mu}\equiv\frac{dx^{\mu}}{ds}=p^{\mu}-\frac{2}{\omega}B^{\mu}\,.
\ee
Then keeping up to order $1/\omega$ terms we can rewrite equation \eq{3.23} as follows
\be\n{3.25}
l^{\nu}l^{\mu}_{\;\;;\nu} =\frac{2}{\omega}K^{\mu}_{\;\;\nu}l^{\nu}\,,
\ee
where we defined
\be\n{3.26}
K_{\mu\nu}\equiv B_{\nu ;\mu}-B_{\mu;\nu}\,.
\ee

Note that equations \eq{3.25} are similar to the equations of motion for a particle with electric charge $2/\omega$ moving in a magnetic field with the vector potential $B_{\mu}$. The tensor $K_{\mu\nu}$ is analogous to the Faraday tensor. For electromagnetic spinoptics the corresponding "charge" is twice less.  This factor 2 difference can be naturally explained in the framework of
the gravitoelectromagnetism (see e.g. \cite{Mashhoon:1999nr} and references therein).

The equation \eqref{3.25} implies that
\be\n{3.27}
l^{\mu}(l^{\nu}l_{\nu})_{;\mu}=0\,.
\ee
This equation shows that the norm of the tangent vector $l^{\mu}$ is preserved along the ray. According to the expression \eq{3.16}, this norm vanishes. Thus the vector and the ray are null.

\subsection{Complex null tetrads associated with null rays }

In order to formulate spinoptics equations we use a special complex null tetrad $(l^{\mu},m^{\mu},\bar{m}^{\mu},n^{\mu})$ associated with the congruence of null rays \eq{3.24} and normalised as follows:
\be\n{3.28}
m^{\mu}\bar{m}_{\mu}=-l^{\mu}n_{\mu}=1\,,
\ee
where all other scalar products are zero.

Let us denote $D=l^{\mu}\nabla_{\mu}$. Then using the normalization conditions for the null complex tetrad one can write the following equation for the propagation of the tetrad along $\BM{l}$,
\ba\n{3.29}
\begin{split}
& D\BM{l}&=&(\varepsilon+\bar{\varepsilon})\BM{l}-\bar{\kappa}_1\BM{m}-\kappa_1\bar{\BM{m}}\,,\\
& D\BM{n}&=&-(\varepsilon+\bar{\varepsilon})\BM{n}+\pi\BM{m}+\bar{\pi}\bar{\BM{m}}\,,\\
& D\BM{m}&=&(\varepsilon-\bar{\varepsilon})\BM{m}+\bar{\pi}\BM{l}-\kappa\BM{n}\,,
\end{split}
\ea
Here $\varepsilon$, $\kappa_1$ and $\pi$ are standard Newman-Penrose coefficients\footnote{As we shall see, the standard NP coefficient $\kappa$ is of order of $1/\omega$. To stress this, we denote $\kappa$ by $\kappa_1$.} (see for example \cite{Chandra}).

We assume that a null tetrad in limit $\omega\to \infty$ becomes parallel propagated along $\BM{l}$.
Let us note that there is an ambiguity in the choice of the complex null tetrad. Namely, the following transformations preserve the direction of the null vector $\BM{l}$ and the normalization conditions \eqref{3.29}
\be
\begin{split}
& (i) \;\;\;\BM{l}\to\BM{l}\hh\BM{m}\to \BM{m}+a\BM{l}\hh\bar{\BM{m}}\to\bar{\BM{m}}+\bar{a}\BM{l}\,,\non\\
& \qquad\BM{n}\to\BM{n}+\bar{a}\BM{m}+a\bar{\BM{m}}+a\bar{a}\BM{l}\,;\\
& (ii) \;\;\;\BM{l}\to\gamma\BM{l}\hh\BM{n}\to\gamma^{-1}\BM{n}\, ,\\
&\qquad \BM{m}\to e^{i\phi}\BM{m}\hh
\bar{\BM{m}}\to e^{-i\phi}\bar{\BM{m}}\,.\non
\end{split}
\ee
Under the type {\it{(i)}} transformation the  spin coefficients in \eq{3.29} change as follows:
\be\n{3.31}
\begin{split}
&\kappa_1\to\kappa_1\hh\varepsilon\to\varepsilon+\bar{a}\kappa_1\, ,\\
&\pi\to\pi+2\bar{a}\varepsilon+\bar{a}^2\kappa_1+D\bar{a}\,.
\end{split}
\ee
Thus by taking an appropriate $a$ one can put $\pi= 0$.

Under the type {\it{(ii)}} transformations the  spin coefficients in \eq{3.29} change as follows:
\be\n{3.32}
\begin{split}
&\kappa_1\to\gamma^2e^{i\phi}\kappa_1
\hh\pi\to\pi e^{-i\phi}\, ,\\
&\varepsilon\to\gamma\varepsilon+\frac{1}{2}D\gamma+\frac{i}{2}\gamma D\phi\,  .
\end{split}
\ee
One can take $\gamma$ and $\phi$ such that $\varepsilon = 0$.  In what follows we impose these conditions $\varepsilon =\pi = 0$.

A comparison of \eq{3.25} with the propagation equation for $\BM{l}$ in \eq{3.29} shows that
\be
\n{3.34}
K^{\mu}_{\;\;\nu}l^{\nu}=\bar{\kappa}m^{\mu}+{\kappa}\bar{m}^{\mu}\hh \kappa_1=-2\kappa/\omega\, .
\ee
Then the equations \eqref{3.29} take the form
\be\n{3.33}
Dl^{\mu}=\frac{2}{\omega}(\bar{\kappa}m^{\mu}+{\kappa}\bar{m}^{\mu})\hh Dn^{\mu}=0\hh Dm^{\mu}=\frac{2}{\omega}\kappa n^{\mu}\, .
\ee

\subsection{Using complex null tetrads associated with the null rays}

Our next step is to write the equations \eq{3.17}, \eq{3.20}, \eqref{3.25} in terms of the congruence of null rays and the complex null tetrad associated with it. We present the complex null vector $\BM{M}$ in the form
\be\n{3.35}
M^{\alpha}=m^{\alpha}+\frac{2}{\omega}\mu^{\alpha}\,,
\ee
where $\mu^{\alpha}$ is a complex vector of the order of 1. Then using the conditions \eq{MMM} and keeping terms up to order $1/\omega$ we get
\be\n{3.36}
\mu^{\alpha}m_{\alpha}=0\hh \bar{\mu}^{\alpha}m_{\alpha}+\mu^{\alpha}\bar{m}_{\alpha}=0\,.
\ee
Similarly, using the definition \eq{3.11} of the vector $B^{\mu}$ and neglecting the terms of order $1/\omega^2$ we derive
\ba\n{3.37}
B^{\mu}&=&b^{\mu}+\frac{2}{\omega}\beta^{\mu}\hh b_{\mu}=i\bar{m}^{\nu}m_{\nu;\mu}\,,\non\\
\beta_{\alpha}&=&i(\bar{\mu}^{\nu}m_{\nu;\alpha}+\bar{m}^{\nu}\mu_{\nu;\alpha})\,.
\ea
Since the difference between $B_{\mu}$ and $b_{\mu}$ is of order $1/\omega$, we can replace $B_{\mu}$ by $b_{\mu}$ in equations \eq{3.24}, \eq{3.25} and \eq{3.26}. Keeping the terms up to the first order in $1/\omega$ one gets
\be\n{3.38}
\begin{split}
&l^{\mu}=p^{\mu}-\frac{2}{\omega}b^{\mu}\, ,\\
&l^{\nu}l^{\mu}_{\;\;;\nu}=\frac{2}{\omega}k^{\mu}_{\;\;\nu}l^{\nu}\hh k_{\mu\nu}=b_{\nu;\mu}-b_{\mu;\nu}\,.
\end{split}
\ee
Using the explicit form of $b_{\mu}$ in \eq{3.37} we derive
\ba\n{3.40}
k_{\mu\nu}l^{\nu}&=&i\bar{m}^{\alpha}(m_{\alpha;\nu\mu}-m_{\alpha;\mu\nu})l^{\nu}\non\\
&+&i (\bar{m}_{\alpha ;\mu}Dm^{\alpha}-{m}_{\alpha ;\mu}D\bar{m}^{\alpha})\,.
\ea
The propagation equation \eq{3.33} implies that  the second term on the r.h.s of this relation is of order $1/\omega$ and therefore it can be neglected. Expressing commutator of the covariant derivatives in \eq{3.40} via the Riemann curvature tensor we write the propagation equation in \eq{3.38} as follows:
\be\n{3.41}
l^{\nu}l^{\mu}_{\;\;;\nu}=\frac{2i}{\omega}R^{\mu}_{\;\;\nu\alpha\beta}l^{\nu}\bar{m}^{\alpha}m^{\beta}\,.
\ee

Let us now consider equation \eq{3.20} with $S_{;\mu}=p_{\mu}$. Using the last equation in \eq{3.33}, \eq{3.35}, and \eq{3.38} and neglecting terms of the order $1/\omega^3$ we derive
\be\n{3.42}
\frac{4i}{\omega^2}A^2(D\mu^{\alpha}+\kappa n^{\alpha}+b^{\beta}m^{\alpha}_{\ ;\beta})=\lambda_3\left(l^{\alpha}+\frac{2}{\omega}b^{\alpha}\right)\, .
\ee
This expression implies that $\lambda_3$ is of order $1/\omega^2$. Therefore, the terms with the Lagrange multiplier $\lambda_3$ in the current \eq{3.18} can be omitted and the continuity equation \eq{3.17} takes the form
\be\n{3.43}
\left(A^2l^{\mu}\right)_{;\mu}=0\,.
\ee
This equation implies that the number of ``gravitons" in a beam formed by null rays is conserved.

Finally, let us consider the last relation in \eqref{MMM}, that is, $M^{\mu}p_{\mu}=0$. Using \eq{3.35} and \eq{3.38}, this relation in the leading order is satisfied, while in the sub-leading order it gives
\be\n{3.44}
l^{\alpha}\mu_{\alpha}=-m^{\alpha}b_{\alpha}\,.
\ee
Note that the relations \eq{3.36} and \eq{3.44} do not specify $\mu^{\alpha}$ uniquely. Because the condition \eq{3.15} implies that $l^{\mu}b_{\mu}$ is of order $1/\omega$, they remain valid under the transformation
\be\n{3.45}
\mu^{\alpha}\to\mu^{\alpha}+\nu l^{\alpha}\,,
\ee
with the gauge scalar function $\nu$. Denote now $\lambda_3=4i\lambda A^2/\omega^2$.
Then in the leading order equation \eq{3.42} takes
the form
\be\n{3.47}
D\mu^{\alpha}+\kappa n^{\alpha}+b^{\beta}m^{\alpha}_{\;\;;\beta}=\lambda l^{\alpha}\, .
\ee
Using the gauge freedom \eq{3.45} we can put $\lambda=0$, that implies the propagation law
\be\n{3.48}
D\mu^{\alpha}=-\kappa n^{\alpha}-b^{\beta}m^{\alpha}_{\;\;;\beta}\,.
\ee

\subsection{General form of the spinoptics equations}

The above effective action and spinoptics equations were derived for the special ansatz \eqref{3.1} of the complex perturbation field $h_{\mu\nu}$ which corresponds to the choice of the helicity $\sigma=+2$. By changing $m_{\mu}\to \bar{m}_{\mu}$ and repeating the calculations one can easily obtain the spinoptics equation for the negative value of the helicity. It is easy to check that the following quantities are sensitive to the sign of the helicity and under the change $m_{\mu}\to \bar{m}_{\mu}$ they transform as
\be\n{3.50}
\begin{split}
&b_{\mu}\to - b_{\mu}\hh k_{\mu\nu}\to - k_{\mu\nu}\hh \kappa\to -\kappa\,.
\end{split}
\ee

Using this observation one can easily present the  gravitational spinoptics equations in a form, valid for both signs of the helicity
\begin{itemize}
\item The null ray equation
\be\n{3.51a}
l^{\nu}l^{\mu}_{\;\;\;;\nu}=\frac{i\sigma}{\omega}R^{\mu}_{\;\;\nu\alpha\beta}l^{\nu}\bar{m}^{\alpha}m^{\beta}\,.
\ee
\item The null tetrad propagation equations
\be\n{3.51b}
\begin{split}
& Dn^{\mu}=0\hh Dm^{\mu}=\frac{\sigma}{\omega}\kappa n^{\mu}\,,\\
&\kappa=iR_{\mu\nu\alpha\beta}m^{\mu}l^{\nu}\bar{m}^{\alpha}m^{\beta}\,.
\end{split}
\ee
\item The continuity equation
\be\n{3.51d}
\left(A^2l^{\mu}\right)_{;\mu}=0\,.
\ee
\item Polarization correction equation
\be\n{3.51e}
D\mu^{\alpha}=-\sigma\left(\kappa n^{\alpha}+b^{\beta}m^{\alpha}_{\;\beta}\right)\hh
b_{\mu}=i\bar{m}^{\nu}m_{\nu;\mu}\,.
\ee
\end{itemize}

Note that we use $\sigma$ for the helicity and it takes values $\pm 2$ for gravitons and $\pm 1$ for photons.

\section{Discussion}

In this work we extended the spinoptics formalism for Maxwell equations in a curved spacetime, developed earlier in \cite{Frolov:2024ebe}, to the case of high-frequency gravitational waves. This approach is based on the calculation of the effective action which is a functional of the wave amplitude $A$, its phase $S$, and a complex null vector $M^{\mu}$.
This derivation of the spinoptics equations is more transparent as compared with the earlier derivations of the spinoptical equations presented in \cite{Oancea:2019pgm,Oancea:2020khc,Frolov:2020uhn}.

A solution of the massless field equations in the high-frequency approximation reduces to study a set of zero-mass particle trajectories. Electromagnetic field and gravitational waves possess spin, and this property makes them different from the scalar massless field. Photons and gravitons have helicity and it should be included in the description of their dynamics. For particles propagating in the external gravitational field their spin interacts with the spacetime curvature. This interaction modifies dynamics of photons and gravitons with respect to the massless scalar particles.  As a result, the particles, associated with the waves of different polarisation (helicity) propagate along different paths. In the spinoptics equation \eq{3.51a} this difference is due to the ``force term" on the r.h.s. which is proportional to the helicity.  This implies that under the same conditions within the spinoptics formalism the helicity induced separation between the right- and left-handed gravitational waves is twice larger than for the electromagnetic waves.

Let us emphasized that the  gravitational spinoptics equations \eqref{3.51a}--\eqref{3.51e} formally coincide with similar equations for the electromagnetic spinoptics \cite{Frolov:2024ebe}. The only difference is that for the case of gravity the helicity parameter $\sigma$ take values $\pm 2$, while for the Maxwell equations $\sigma=\pm 1$.
If one put $\sigma=0$ in these equations all the dependence on the helicity and spin disappears. In this limit the obtained equations reduce to the standard equations for the geometric optics approximation, where the null rays are null geodesics and the tetrad associated with the rays is parallel transported along the rays.

The system of equations \eqref{3.51a}--\eqref{3.51e} can be used to find a solution for high-frequency gravitational waves propagating in a curved spacetime in the framework of the  spinoptics approximation. For this purpose one needs to specify an initial condition for the high-frequency gravitational wave at some initial timelike surface $\Sigma$. These conditions determine the initial vectors $l^{\mu}$ of the null rays at $\Sigma$, as well the initial value of
the complex null tetrad associated with them. Solution of the
equations \eqref{3.51a}--\eqref{3.51e} determines the null rays and complex null tetrads outside $\Sigma$. To find the eikonal function $S$ one uses the following relation
for the change of $S$ along the null rays
\be \n{SSS}
\int dx^{\mu}S_{,\mu}=\int ds l^{\mu} (l_{\mu}+\dfrac{q}{\omega}b_{\mu})=O(1/\omega^2)\, .
\ee
The integration is performed along corresponding null rays.
Equation \eqref{SSS} shows that in the high-frequency approximation the change of $S$ along the null rays is of
the second order in $1/\omega$. That is in an adopted approximation  the phase function (eikonal) is constant along the null rays. If in some vicinity of $\Sigma$ the null rays of the congruence do not intersect, and there is no caustics, the value of $S$ at a point $p$ can be found as follows. Consider a null ray passing through this point and trace it back in time until it crosses the initial surface $\Sigma$ at some point $p_0$. The value $S(p)$ coincides with the initial value $S_0$ of the phase function $S$ at $p_0$ on $\Sigma$ , $S(p)=S_0(p_0)$

The method of the effective action for the spinoptics of the electromagnetic and gravitational waves is totally covariant and it can be applied for an arbitrary vacuum spacetime background. One can expect that in the presence of explicit or/and hidden symmetries of the background geometry the spinoptics equations could be simplified and even, under special conditions, become completely integrable. It is interesting to investigate this in more details. It is also interesting to generalize the developed in this paper approach to the case  of the gravitational high-frequency waves  propagating in the spacetime filled with the matter.
It might be also interesting to extend the presented spinoptics formalism and to include higher order terms in  the $1/\omega$ expansion.

\appendix

\section{Second order perturbation of the Einstein-Hilbert action}

Here we briefly summarize a derivation of a second order perturbation of the Einstein-Hilbert action. Assuming that  $|h_{\mu\nu}|\ll|g_{\mu\nu}|$ in the chosen coordinate frame $x^{\mu}$, we write
\be\n{A1}
\hat{g}_{\mu\nu}=g_{\mu\nu}+h_{\mu\nu}\, ,
\ee
where $\hat{g}_{\mu\nu}$ is the spacetime metric, $g_{\mu\nu}$ is the background spacetime, and $h_{\mu\nu}$ is its perturbation.
Then expanding different geometric objects in powers of $h_{\mu\nu}$ and keeping the terms up to the second order one has
\ba\n{A2}
\begin{split}
&\hat{g}^{\mu\nu} = g^{\mu\nu}-h^{\mu\nu}+h^{\mu\alpha}h_{\alpha}^{\;\;\nu}\, ,\\
&\hat{g}=g\left(1+h+\frac{1}{2}h^2-\frac{1}{2}h^{\mu\nu}h_{\mu\nu}\right)\,,\non\\
&\sqrt{-\hat{g}}=\sqrt{-g}\left(1+\frac{1}{2}h+\frac{1}{8}h^2-\frac{1}{4}h^{\mu\nu}h_{\mu\nu}\right)\,,
\end{split}
\ea
where
\be\n{A3}
\hat{g}=\det(\hat{g}_{\mu\nu})\hh g=\det(g_{\mu\nu})\hh h=h^{\mu}_{\;\mu}\,.
\ee
Here and in what follows, the indices are raised and lowered with $g_{\mu\nu}$. Using these expressions one can construct second order correction to the space-time Ricci tensor,
\be\n{A4}
\hat{R}_{\mu\nu}=R_{\mu\nu}+R^{(1)}_{\mu\nu}+R^{(2)}_{\mu\nu}\,,
\ee
where $R_{\mu\nu}$ is the background Ricci tensor and [cf. p. 965 \cite{MTW}]
\ba\n{A5}
R^{(1)}_{\mu\nu}&=&\frac{1}{2}\left(h^{\alpha}_{\;\mu;\nu\alpha}+h^{\alpha}_{\;\nu;\mu\alpha}-h_{\mu\nu;\alpha}^{\quad\;\;\;;\alpha}-h_{;\mu\nu}\right)\,,\non\\
R^{(2)}_{\mu\nu}&=&\frac{1}{2}\left(\frac{1}{2}h_{\alpha\beta;\mu}h^{\alpha\beta}_{\;\;\;\;;\nu}+h_{\nu}^{\;\;\alpha;\beta}(h_{\alpha\mu;\beta}-h_{\beta\mu;\alpha})\right.\\
&+&h^{\alpha\beta}(h_{\alpha\beta;\mu\nu}+h_{\mu\nu;\alpha\beta}-h_{\alpha\mu;\nu\beta}-h_{\alpha\nu;\mu\beta})\non\\
&-&\left.\left[h^{\alpha\beta}_{\;\;\;\;;\beta}-\frac{1}{2}h^{;\alpha}\right](h_{\alpha\mu;\nu}+h_{\alpha\nu;\mu}-h_{\mu\nu;\alpha})\right]\, ,\non
\ea
where the semicolon stands for the covariant derivative associated with the background metric.

Consider now the Einstein-Hilbert action for the gravitational field,
\be\n{A6}
\hat{I}=\frac{1}{16\pi}\int d^4x\sqrt{-\hat{g}}\,\hat{g}^{\mu\nu}\hat{R}_{\mu\nu}
\ee
Plugging \eqref{A1}-- \eqref{A5} into the action \eq{A6} and expanding it in powers of $h_{\mu\nu}$ up to second order we get
\be\n{A7}
\hat{I}=I^{(0)}+I^{(1)}+I^{(2)}\,,
\ee
where
\ba\n{A8}
I^{(0)}&=&\frac{1}{16\pi}\int d^4x\sqrt{-g}\,g^{\mu\nu}R_{\mu\nu}\,,\non\\
I^{(1)}&=&\frac{1}{16\pi}\int d^4x\sqrt{-g}\,\left(g^{\mu\nu}R^{(1)}_{\mu\nu}-h^{\mu\nu}G_{\mu\nu}\right)\,,\\
I^{(2)}&=&\frac{1}{16\pi}\int d^4x\sqrt{-g}\,\left(g^{\mu\nu}R^{(2)}_{\mu\nu}-h^{\mu\nu}G^{(1)}_{\mu\nu}\right.\non\\
&+&\left.\left[h^{\mu\alpha}h_{\alpha}^{\;\;\nu}-\frac{1}{2}hh^{\mu\nu}\right]G_{\mu\nu}+\frac{1}{4}\left[h^{\mu\nu}h_{\mu\nu}-\frac{1}{2}h^{2}\right]R\right)\,.\non
\ea
Here
\be\n{A9}
\begin{split}
& G_{\mu\nu}=R_{\mu\nu}-\frac{1}{2}g_{\mu\nu}R\, ,\\ &G^{(1)}_{\mu\nu}=R^{(1)}_{\mu\nu}-\frac{1}{2}g_{\mu\nu}g^{\alpha\beta}R^{(1)}_{\alpha\beta}\,.
\end{split}
\ee
In vacuum spacetimes ($G_{\mu\nu}=0$), after integrating by parts and ignoring vanishing boundary terms, the second order action reduces to
\ba\n{A10}
I_2&=&-\frac{1}{32\pi}\int d^4x\sqrt{-g}\left(\frac{1}{2}h_{\mu\nu;\alpha}h^{\mu\nu;\alpha}-h_{\mu\nu;\alpha}h^{\alpha\nu;\mu}\right.\non\\
&+&\left.h^{\mu\nu}_{\,\,\,\,\,\,;\nu}h_{;\mu}-\frac{1}{2}h_{;\mu}h^{;\mu}\right)\,.
\ea
The propagation equation for gravitational waves, $R^{(1)}_{\mu\nu}=0$, can be derived by extremizing the action with respect to $h^{\mu\nu}$, that gives
\be\n{A11}
h_{\mu\nu;\alpha}^{\quad\;\;\;;\alpha}-g_{\mu\nu}h_{;\alpha}^{\;\;\;;\alpha}+g_{\mu\nu}h^{\alpha\beta}_{\;\;\;\;;\alpha\beta}+h_{;\mu\nu}-h_{\alpha\mu;\nu}^{\quad\;\;\;;\alpha}-h_{\alpha\nu;\mu}^{\quad\;\;\;;\alpha}=0\,.
\ee
Contracting this equation with $g^{\mu\nu}$ gives
\be\n{A12}
h^{\mu\nu}_{\;\;\;\;;\mu\nu}-h_{;\mu}^{\;\;\;;\mu}=0\,,
\ee
and subtracting \eq {A12} multiplied by $g_{\mu\nu}$ from \eq{A11} gives the propagation equation
\be\n{A13}
h_{\mu\nu;\alpha}^{\quad\;\;\;;\alpha}+h_{;\mu\nu}-h_{\alpha\mu;\nu}^{\quad\;\;\;;\alpha}-h_{\alpha\nu;\mu}^{\quad\;\;\;;\alpha}=0\,.
\ee

\vspace{0.5cm}

\section*{Acknowledgments}

One of the authors (V.F.) thanks the Natural Sciences and Engineering Research Council of Canada and the Killam Trust for their financial support. The other author (A.S.)  was supported by the Ministry of Science and Higher Education of the Russian Federation (agreement no. 075-02-2024-1430).

%\bibliography{SPIN_2}

\begin{thebibliography}{46}%
\makeatletter
\providecommand \@ifxundefined [1]{%
 \@ifx{#1\undefined}
}%
\providecommand \@ifnum [1]{%
 \ifnum #1\expandafter \@firstoftwo
 \else \expandafter \@secondoftwo
 \fi
}%
\providecommand \@ifx [1]{%
 \ifx #1\expandafter \@firstoftwo
 \else \expandafter \@secondoftwo
 \fi
}%
\providecommand \natexlab [1]{#1}%
\providecommand \enquote  [1]{``#1''}%
\providecommand \bibnamefont  [1]{#1}%
\providecommand \bibfnamefont [1]{#1}%
\providecommand \citenamefont [1]{#1}%
\providecommand \href@noop [0]{\@secondoftwo}%
\providecommand \href [0]{\begingroup \@sanitize@url \@href}%
\providecommand \@href[1]{\@@startlink{#1}\@@href}%
\providecommand \@@href[1]{\endgroup#1\@@endlink}%
\providecommand \@sanitize@url [0]{\catcode `\\12\catcode `\$12\catcode
  `\&12\catcode `\#12\catcode `\^12\catcode `\_12\catcode `\%12\relax}%
\providecommand \@@startlink[1]{}%
\providecommand \@@endlink[0]{}%
\providecommand \url  [0]{\begingroup\@sanitize@url \@url }%
\providecommand \@url [1]{\endgroup\@href {#1}{\urlprefix }}%
\providecommand \urlprefix  [0]{URL }%
\providecommand \Eprint [0]{\href }%
\providecommand \doibase [0]{http://dx.doi.org/}%
\providecommand \selectlanguage [0]{\@gobble}%
\providecommand \bibinfo  [0]{\@secondoftwo}%
\providecommand \bibfield  [0]{\@secondoftwo}%
\providecommand \translation [1]{[#1]}%
\providecommand \BibitemOpen [0]{}%
\providecommand \bibitemStop [0]{}%
\providecommand \bibitemNoStop [0]{.\EOS\space}%
\providecommand \EOS [0]{\spacefactor3000\relax}%
\providecommand \BibitemShut  [1]{\csname bibitem#1\endcsname}%
\let\auto@bib@innerbib\@empty
%</preamble>
\bibitem [{\citenamefont {Abbott}\ and\ \citenamefont {et.
  al.}(2016)}]{PhysRevLett.116.061102}%
  \BibitemOpen
  \bibfield  {author} {\bibinfo {author} {\bibfnamefont {B.~P.}\ \bibnamefont
  {Abbott}}\ and\ \bibinfo {author} {\bibnamefont {et. al.}} (\bibinfo
  {collaboration} {LIGO Scientific Collaboration and Virgo Collaboration}),\
  }\bibfield  {title} {\enquote {\bibinfo {title} {Observation of gravitational
  waves from a binary black hole merger},}\ }\href {\doibase
  10.1103/PhysRevLett.116.061102} {\bibfield  {journal} {\bibinfo  {journal}
  {Phys. Rev. Lett.}\ }\textbf {\bibinfo {volume} {116}},\ \bibinfo {pages}
  {061102} (\bibinfo {year} {2016})}\BibitemShut {NoStop}%
\bibitem [{\citenamefont {Misner}\ \emph {et~al.}(1974)\citenamefont {Misner},
  \citenamefont {Thorne},\ and\ \citenamefont {Wheeler}}]{MTW}%
  \BibitemOpen
  \bibfield  {author} {\bibinfo {author} {\bibfnamefont {Charles~W.}\
  \bibnamefont {Misner}}, \bibinfo {author} {\bibfnamefont {K.S.}\ \bibnamefont
  {Thorne}}, \ and\ \bibinfo {author} {\bibfnamefont {J.A.}\ \bibnamefont
  {Wheeler}},\ }\href@noop {} {\emph {\bibinfo {title} {{Gravitation}}}}\
  (\bibinfo  {publisher} {W.H. Freeman and Co., San Francisco},\ \bibinfo
  {year} {1974})\BibitemShut {NoStop}%
\bibitem [{\citenamefont {Skrotskii}(1957)}]{Skrotskii}%
  \BibitemOpen
  \bibfield  {author} {\bibinfo {author} {\bibfnamefont {G.~V.}\ \bibnamefont
  {Skrotskii}},\ }\bibfield  {title} {\enquote {\bibinfo {title} {{On the
  Influence of Gravity on the Light Propagation}},}\ }\href@noop {} {\bibfield
  {journal} {\bibinfo  {journal} {Soviet Phys. Doklady}\ }\textbf {\bibinfo
  {volume} {2}},\ \bibinfo {pages} {226} (\bibinfo {year} {1957})},\ \bibinfo
  {note} {[Akademia Nauk SSR, Doklady, {\bf 114}, 73, 1957]}\BibitemShut
  {NoStop}%
\bibitem [{\citenamefont {Plebanski}(1959)}]{Plebanski}%
  \BibitemOpen
  \bibfield  {author} {\bibinfo {author} {\bibfnamefont {J.}~\bibnamefont
  {Plebanski}},\ }\bibfield  {title} {\enquote {\bibinfo {title}
  {{Electromagnetic Waves in Gravitational Fields}},}\ }\href@noop {}
  {\bibfield  {journal} {\bibinfo  {journal} {Phys. Rev.}\ }\textbf {\bibinfo
  {volume} {118}},\ \bibinfo {pages} {1396} (\bibinfo {year}
  {1959})}\BibitemShut {NoStop}%
\bibitem [{\citenamefont {Godfrey}(1970)}]{God}%
  \BibitemOpen
  \bibfield  {author} {\bibinfo {author} {\bibfnamefont {B.~B.}\ \bibnamefont
  {Godfrey}},\ }\bibfield  {title} {\enquote {\bibinfo {title} {{Mach's
  Principle, the Kerr Metric, and Black-Hole Physics}},}\ }\href@noop {}
  {\bibfield  {journal} {\bibinfo  {journal} {Phys. Rev. D}\ }\textbf {\bibinfo
  {volume} {1}},\ \bibinfo {pages} {2721} (\bibinfo {year} {1970})}\BibitemShut
  {NoStop}%
\bibitem [{\citenamefont {Pineault}\ and\ \citenamefont {Roeder}(1977)}]{GF2}%
  \BibitemOpen
  \bibfield  {author} {\bibinfo {author} {\bibfnamefont {S.}~\bibnamefont
  {Pineault}}\ and\ \bibinfo {author} {\bibfnamefont {R.~C.}\ \bibnamefont
  {Roeder}},\ }\bibfield  {title} {\enquote {\bibinfo {title} {{Applications of
  Geometrical Optics to the Kerr Metric. I. Analytical Results}},}\ }\href@noop
  {} {\bibfield  {journal} {\bibinfo  {journal} {Astrophys. J.}\ }\textbf
  {\bibinfo {volume} {212}},\ \bibinfo {pages} {541} (\bibinfo {year}
  {1977})}\BibitemShut {NoStop}%
\bibitem [{\citenamefont {Connors}\ and\ \citenamefont {Stark}(1977)}]{GF3}%
  \BibitemOpen
  \bibfield  {author} {\bibinfo {author} {\bibfnamefont {P.~A.}\ \bibnamefont
  {Connors}}\ and\ \bibinfo {author} {\bibfnamefont {R.~F.}\ \bibnamefont
  {Stark}},\ }\bibfield  {title} {\enquote {\bibinfo {title} {{Observable
  gravitational effects on polarised radiation coming from near a black
  hole}},}\ }\href@noop {} {\bibfield  {journal} {\bibinfo  {journal} {Nature
  (London)}\ }\textbf {\bibinfo {volume} {269}},\ \bibinfo {pages} {128}
  (\bibinfo {year} {1977})}\BibitemShut {NoStop}%
\bibitem [{\citenamefont {Connors}\ \emph {et~al.}(1980)\citenamefont
  {Connors}, \citenamefont {Piran},\ and\ \citenamefont {Stark}}]{GF4}%
  \BibitemOpen
  \bibfield  {author} {\bibinfo {author} {\bibfnamefont {P.~A.}\ \bibnamefont
  {Connors}}, \bibinfo {author} {\bibfnamefont {T.}~\bibnamefont {Piran}}, \
  and\ \bibinfo {author} {\bibfnamefont {R.~F.}\ \bibnamefont {Stark}},\
  }\bibfield  {title} {\enquote {\bibinfo {title} {{Polarization features of
  X-ray radiation emitted near black holes}},}\ }\href@noop {} {\bibfield
  {journal} {\bibinfo  {journal} {Astrophys. J.}\ }\textbf {\bibinfo {volume}
  {235}},\ \bibinfo {pages} {224} (\bibinfo {year} {1980})}\BibitemShut
  {NoStop}%
\bibitem [{\citenamefont {Fayos}\ and\ \citenamefont {Llosa}(1982)}]{GF5}%
  \BibitemOpen
  \bibfield  {author} {\bibinfo {author} {\bibfnamefont {F.}~\bibnamefont
  {Fayos}}\ and\ \bibinfo {author} {\bibfnamefont {J.}~\bibnamefont {Llosa}},\
  }\bibfield  {title} {\enquote {\bibinfo {title} {{Gravitational Effects on
  the Polarization Plane}},}\ }\href@noop {} {\bibfield  {journal} {\bibinfo
  {journal} {General Relativity and Gravitation}\ }\textbf {\bibinfo {volume}
  {14}},\ \bibinfo {pages} {865} (\bibinfo {year} {1982})}\BibitemShut
  {NoStop}%
\bibitem [{\citenamefont {{Piran}}\ and\ \citenamefont
  {{Safier}}(1985)}]{Piran:1985}%
  \BibitemOpen
  \bibfield  {author} {\bibinfo {author} {\bibfnamefont {T.}~\bibnamefont
  {{Piran}}}\ and\ \bibinfo {author} {\bibfnamefont {P.~N.}\ \bibnamefont
  {{Safier}}},\ }\bibfield  {title} {\enquote {\bibinfo {title} {{A
  gravitational analogue of Faraday rotation}},}\ }\href@noop {} {\bibfield
  {journal} {\bibinfo  {journal} {\nat}\ }\textbf {\bibinfo {volume} {318}},\
  \bibinfo {pages} {271--273} (\bibinfo {year} {1985})}\BibitemShut {NoStop}%
\bibitem [{\citenamefont {Ishihara}\ \emph {et~al.}(1988)\citenamefont
  {Ishihara}, \citenamefont {Takahashi},\ and\ \citenamefont
  {Tomimatsu}}]{GF6}%
  \BibitemOpen
  \bibfield  {author} {\bibinfo {author} {\bibfnamefont {H.}~\bibnamefont
  {Ishihara}}, \bibinfo {author} {\bibfnamefont {M.}~\bibnamefont {Takahashi}},
  \ and\ \bibinfo {author} {\bibfnamefont {A.}~\bibnamefont {Tomimatsu}},\
  }\bibfield  {title} {\enquote {\bibinfo {title} {{Gravitational Faraday
  rotation induced by a Kerr black hole}},}\ }\href@noop {} {\bibfield
  {journal} {\bibinfo  {journal} {Phys.\ Rev.\ D}\ }\textbf {\bibinfo {volume}
  {38}},\ \bibinfo {pages} {472} (\bibinfo {year} {1988})}\BibitemShut
  {NoStop}%
\bibitem [{\citenamefont {Carini}\ \emph {et~al.}(1992)\citenamefont {Carini},
  \citenamefont {Feng}, \citenamefont {Li},\ and\ \citenamefont
  {Ruffini}}]{CariniRuffini}%
  \BibitemOpen
  \bibfield  {author} {\bibinfo {author} {\bibfnamefont {P.}~\bibnamefont
  {Carini}}, \bibinfo {author} {\bibfnamefont {L.~L.}\ \bibnamefont {Feng}},
  \bibinfo {author} {\bibfnamefont {M.}~\bibnamefont {Li}}, \ and\ \bibinfo
  {author} {\bibfnamefont {R.}~\bibnamefont {Ruffini}},\ }\bibfield  {title}
  {\enquote {\bibinfo {title} {{Phase evolution of the photon in Kerr
  spacetime}},}\ }\href@noop {} {\bibfield  {journal} {\bibinfo  {journal}
  {Phys.\ Rev.\ D}\ }\textbf {\bibinfo {volume} {46}},\ \bibinfo {pages} {5407}
  (\bibinfo {year} {1992})}\BibitemShut {NoStop}%
\bibitem [{\citenamefont {Perlick}\ and\ \citenamefont
  {Hasse}(1993)}]{Perlick_1993}%
  \BibitemOpen
  \bibfield  {author} {\bibinfo {author} {\bibfnamefont {V.}~\bibnamefont
  {Perlick}}\ and\ \bibinfo {author} {\bibfnamefont {W.}~\bibnamefont
  {Hasse}},\ }\bibfield  {title} {\enquote {\bibinfo {title} {{Gravitational
  Faraday effect in conformally stationary spacetimes}},}\ }\href {\doibase
  10.1088/0264-9381/10/1/015} {\bibfield  {journal} {\bibinfo  {journal}
  {Classical and Quantum Gravity}\ }\textbf {\bibinfo {volume} {10}},\ \bibinfo
  {pages} {147--161} (\bibinfo {year} {1993})}\BibitemShut {NoStop}%
\bibitem [{\citenamefont {Nouri-Zonoz}(1999)}]{GF7}%
  \BibitemOpen
  \bibfield  {author} {\bibinfo {author} {\bibfnamefont {M.}~\bibnamefont
  {Nouri-Zonoz}},\ }\bibfield  {title} {\enquote {\bibinfo {title}
  {{Gravitoelectromagnetic approach to the gravitational Faraday rotation in
  stationary space-times}},}\ }\href@noop {} {\bibfield  {journal} {\bibinfo
  {journal} {Phys.\ Rev.\ D}\ }\textbf {\bibinfo {volume} {60}},\ \bibinfo
  {pages} {024013} (\bibinfo {year} {1999})}\BibitemShut {NoStop}%
\bibitem [{\citenamefont {Sereno}(2004)}]{GF8}%
  \BibitemOpen
  \bibfield  {author} {\bibinfo {author} {\bibfnamefont {M.}~\bibnamefont
  {Sereno}},\ }\bibfield  {title} {\enquote {\bibinfo {title} {{Gravitational
  Faraday rotation in a weak gravitational field}},}\ }\href@noop {} {\bibfield
   {journal} {\bibinfo  {journal} {Phys.\ Rev.\ D}\ }\textbf {\bibinfo {volume}
  {69}},\ \bibinfo {pages} {087501} (\bibinfo {year} {2004})}\BibitemShut
  {NoStop}%
\bibitem [{\citenamefont {Sereno}(2005)}]{GF9}%
  \BibitemOpen
  \bibfield  {author} {\bibinfo {author} {\bibfnamefont {M.}~\bibnamefont
  {Sereno}},\ }\bibfield  {title} {\enquote {\bibinfo {title} {{Detecting
  gravitomagnetism with rotation of polarization by a gravitational lens}},}\
  }\href@noop {} {\bibfield  {journal} {\bibinfo  {journal} {Mon. Not. R.
  Astron. Soc.}\ }\textbf {\bibinfo {volume} {356}},\ \bibinfo {pages} {381}
  (\bibinfo {year} {2005})}\BibitemShut {NoStop}%
\bibitem [{\citenamefont {Halilsoy}\ and\ \citenamefont
  {Gurtug}(2007)}]{Halilsoy:2006ev}%
  \BibitemOpen
  \bibfield  {author} {\bibinfo {author} {\bibfnamefont {Mustafa}\ \bibnamefont
  {Halilsoy}}\ and\ \bibinfo {author} {\bibfnamefont {Ozay}\ \bibnamefont
  {Gurtug}},\ }\bibfield  {title} {\enquote {\bibinfo {title} {{Search for
  gravitational waves through the electromagnetic Faraday rotation}},}\ }\href
  {\doibase 10.1103/PhysRevD.75.124021} {\bibfield  {journal} {\bibinfo
  {journal} {Phys. Rev. D}\ }\textbf {\bibinfo {volume} {75}},\ \bibinfo
  {pages} {124021} (\bibinfo {year} {2007})},\ \Eprint
  {http://arxiv.org/abs/gr-qc/0612107} {arXiv:gr-qc/0612107} \BibitemShut
  {NoStop}%
\bibitem [{\citenamefont {Brodutch}\ \emph {et~al.}(2011)\citenamefont
  {Brodutch}, \citenamefont {Demarie},\ and\ \citenamefont {Terno}}]{GF10}%
  \BibitemOpen
  \bibfield  {author} {\bibinfo {author} {\bibfnamefont {A.}~\bibnamefont
  {Brodutch}}, \bibinfo {author} {\bibfnamefont {T.~F.}\ \bibnamefont
  {Demarie}}, \ and\ \bibinfo {author} {\bibfnamefont {D.~R.}\ \bibnamefont
  {Terno}},\ }\bibfield  {title} {\enquote {\bibinfo {title} {{Photon
  polarization and geometric phase in general relativity}},}\ }\href@noop {}
  {\bibfield  {journal} {\bibinfo  {journal} {Phys.\ Rev.\ D}\ }\textbf
  {\bibinfo {volume} {84}},\ \bibinfo {pages} {104043} (\bibinfo {year}
  {2011})}\BibitemShut {NoStop}%
\bibitem [{\citenamefont {Frolov}\ and\ \citenamefont
  {Shoom}(2011)}]{Frolov:2011mh}%
  \BibitemOpen
  \bibfield  {author} {\bibinfo {author} {\bibfnamefont {Valeri~P.}\
  \bibnamefont {Frolov}}\ and\ \bibinfo {author} {\bibfnamefont {Andrey~A.}\
  \bibnamefont {Shoom}},\ }\bibfield  {title} {\enquote {\bibinfo {title}
  {{Spinoptics in a stationary spacetime}},}\ }\href@noop {} {\bibfield
  {journal} {\bibinfo  {journal} {Phys. Rev. D}\ }\textbf {\bibinfo {volume}
  {84}},\ \bibinfo {pages} {044026} (\bibinfo {year} {2011})}\BibitemShut
  {NoStop}%
\bibitem [{\citenamefont {Ghosh}\ and\ \citenamefont {Sen}(2016)}]{Ghosh}%
  \BibitemOpen
  \bibfield  {author} {\bibinfo {author} {\bibfnamefont {T.}~\bibnamefont
  {Ghosh}}\ and\ \bibinfo {author} {\bibfnamefont {A.~K.}\ \bibnamefont
  {Sen}},\ }\bibfield  {title} {\enquote {\bibinfo {title} {{The Effect of
  Gravitation on the Polarization State of a Light ray}},}\ }\href@noop {}
  {\bibfield  {journal} {\bibinfo  {journal} {Astrophys.\ J.}\ }\textbf
  {\bibinfo {volume} {833}},\ \bibinfo {pages} {82} (\bibinfo {year}
  {2016})}\BibitemShut {NoStop}%
\bibitem [{\citenamefont {Dolan}(2018{\natexlab{a}})}]{Dolan:2018nzc}%
  \BibitemOpen
  \bibfield  {author} {\bibinfo {author} {\bibfnamefont {Sam~R.}\ \bibnamefont
  {Dolan}},\ }\bibfield  {title} {\enquote {\bibinfo {title} {Geometrical
  optics for scalar, electromagnetic and gravitational waves on curved
  spacetime},}\ }\href@noop {} {\bibfield  {journal} {\bibinfo  {journal}
  {International Journal of Modern Physics D}\ }\textbf {\bibinfo {volume}
  {27}},\ \bibinfo {pages} {1843010} (\bibinfo {year}
  {2018}{\natexlab{a}})}\BibitemShut {NoStop}%
\bibitem [{\citenamefont {Dolan}(2018{\natexlab{b}})}]{Dolan:2018ydp}%
  \BibitemOpen
  \bibfield  {author} {\bibinfo {author} {\bibfnamefont {Sam~R.}\ \bibnamefont
  {Dolan}},\ }\bibfield  {title} {\enquote {\bibinfo {title} {{Higher-order
  geometrical optics for electromagnetic waves on a curved spacetime}},}\
  }\href@noop {} {\  (\bibinfo {year} {2018}{\natexlab{b}})},\ \Eprint
  {http://arxiv.org/abs/1801.02273} {arXiv:1801.02273 [gr-qc]} \BibitemShut
  {NoStop}%
\bibitem [{\citenamefont {Hou}\ \emph {et~al.}(2019)\citenamefont {Hou},
  \citenamefont {Fan},\ and\ \citenamefont {Zhu}}]{Hou:2019wdg}%
  \BibitemOpen
  \bibfield  {author} {\bibinfo {author} {\bibfnamefont {Shaoqi}\ \bibnamefont
  {Hou}}, \bibinfo {author} {\bibfnamefont {Xi-Long}\ \bibnamefont {Fan}}, \
  and\ \bibinfo {author} {\bibfnamefont {Zong-Hong}\ \bibnamefont {Zhu}},\
  }\bibfield  {title} {\enquote {\bibinfo {title} {{Gravitational Lensing of
  Gravitational Waves: Rotation of Polarization Plane}},}\ }\href@noop {}
  {\bibfield  {journal} {\bibinfo  {journal} {Phys. Rev. D}\ }\textbf {\bibinfo
  {volume} {100}},\ \bibinfo {pages} {064028} (\bibinfo {year}
  {2019})}\BibitemShut {NoStop}%
\bibitem [{\citenamefont {Li}\ \emph {et~al.}(2022)\citenamefont {Li},
  \citenamefont {Qiao}, \citenamefont {Zhao},\ and\ \citenamefont
  {Er}}]{Li:2022izh}%
  \BibitemOpen
  \bibfield  {author} {\bibinfo {author} {\bibfnamefont {Zhao}\ \bibnamefont
  {Li}}, \bibinfo {author} {\bibfnamefont {Jin}\ \bibnamefont {Qiao}}, \bibinfo
  {author} {\bibfnamefont {Wen}\ \bibnamefont {Zhao}}, \ and\ \bibinfo {author}
  {\bibfnamefont {Xinzhong}\ \bibnamefont {Er}},\ }\bibfield  {title} {\enquote
  {\bibinfo {title} {{Gravitational Faraday Rotation of gravitational waves by
  a Kerr black hole}},}\ }\href {\doibase 10.1088/1475-7516/2022/10/095}
  {\bibfield  {journal} {\bibinfo  {journal} {JCAP}\ }\textbf {\bibinfo
  {volume} {10}},\ \bibinfo {pages} {095} (\bibinfo {year} {2022})},\ \Eprint
  {http://arxiv.org/abs/2204.10512} {arXiv:2204.10512 [gr-qc]} \BibitemShut
  {NoStop}%
\bibitem [{\citenamefont {Shoom}(2022)}]{Shoom:2022oer}%
  \BibitemOpen
  \bibfield  {author} {\bibinfo {author} {\bibfnamefont {Andrey~A.}\
  \bibnamefont {Shoom}},\ }\bibfield  {title} {\enquote {\bibinfo {title}
  {{Faraday effect of light caused by plane gravitational wave}},}\ }\href@noop
  {} {\  (\bibinfo {year} {2022})},\ \Eprint {http://arxiv.org/abs/2206.08867}
  {arXiv:2206.08867 [gr-qc]} \BibitemShut {NoStop}%
\bibitem [{\citenamefont {Frolov}\ and\ \citenamefont
  {Shoom}(2012)}]{Frolov:2012zn}%
  \BibitemOpen
  \bibfield  {author} {\bibinfo {author} {\bibfnamefont {Valeri~P.}\
  \bibnamefont {Frolov}}\ and\ \bibinfo {author} {\bibfnamefont {Andrey~A.}\
  \bibnamefont {Shoom}},\ }\bibfield  {title} {\enquote {\bibinfo {title}
  {{Scattering of circularly polarized light by a rotating black hole}},}\
  }\href@noop {} {\bibfield  {journal} {\bibinfo  {journal} {Phys. Rev. D}\
  }\textbf {\bibinfo {volume} {86}},\ \bibinfo {pages} {024010} (\bibinfo
  {year} {2012})}\BibitemShut {NoStop}%
\bibitem [{\citenamefont {Yoo}(2012)}]{Yoo:2012vv}%
  \BibitemOpen
  \bibfield  {author} {\bibinfo {author} {\bibfnamefont {Chul-Moon}\
  \bibnamefont {Yoo}},\ }\bibfield  {title} {\enquote {\bibinfo {title} {{Notes
  on Spinoptics in a Stationary Spacetime}},}\ }\href@noop {} {\bibfield
  {journal} {\bibinfo  {journal} {Phys. Rev. D}\ }\textbf {\bibinfo {volume}
  {86}},\ \bibinfo {pages} {084005} (\bibinfo {year} {2012})}\BibitemShut
  {NoStop}%
\bibitem [{\citenamefont {Oancea}\ \emph {et~al.}(2019)\citenamefont {Oancea},
  \citenamefont {Paganini}, \citenamefont {Joudioux},\ and\ \citenamefont
  {Andersson}}]{Oancea:2019pgm}%
  \BibitemOpen
  \bibfield  {author} {\bibinfo {author} {\bibfnamefont {Marius~A.}\
  \bibnamefont {Oancea}}, \bibinfo {author} {\bibfnamefont {Claudio~F.}\
  \bibnamefont {Paganini}}, \bibinfo {author} {\bibfnamefont {Jeremie}\
  \bibnamefont {Joudioux}}, \ and\ \bibinfo {author} {\bibfnamefont {Lars}\
  \bibnamefont {Andersson}},\ }\bibfield  {title} {\enquote {\bibinfo {title}
  {{An overview of the gravitational spin Hall effect}},}\ }\href@noop {} {\
  (\bibinfo {year} {2019})},\ \Eprint {http://arxiv.org/abs/1904.09963}
  {arXiv:1904.09963 [gr-qc]} \BibitemShut {NoStop}%
\bibitem [{\citenamefont {Oancea}\ \emph {et~al.}(2020)\citenamefont {Oancea},
  \citenamefont {Joudioux}, \citenamefont {Dodin}, \citenamefont {Ruiz},
  \citenamefont {Paganini},\ and\ \citenamefont {Andersson}}]{Oancea:2020khc}%
  \BibitemOpen
  \bibfield  {author} {\bibinfo {author} {\bibfnamefont {Marius~A.}\
  \bibnamefont {Oancea}}, \bibinfo {author} {\bibfnamefont {Jeremie}\
  \bibnamefont {Joudioux}}, \bibinfo {author} {\bibfnamefont {I.Y.}\
  \bibnamefont {Dodin}}, \bibinfo {author} {\bibfnamefont {D.E.}\ \bibnamefont
  {Ruiz}}, \bibinfo {author} {\bibfnamefont {Claudio~F.}\ \bibnamefont
  {Paganini}}, \ and\ \bibinfo {author} {\bibfnamefont {Lars}\ \bibnamefont
  {Andersson}},\ }\bibfield  {title} {\enquote {\bibinfo {title} {{The
  gravitational spin Hall effect of light}},}\ }\href@noop {} {\  (\bibinfo
  {year} {2020})},\ \Eprint {http://arxiv.org/abs/2003.04553} {arXiv:2003.04553
  [gr-qc]} \BibitemShut {NoStop}%
\bibitem [{\citenamefont {Frolov}(2020)}]{Frolov:2020uhn}%
  \BibitemOpen
  \bibfield  {author} {\bibinfo {author} {\bibfnamefont {Valeri~P.}\
  \bibnamefont {Frolov}},\ }\bibfield  {title} {\enquote {\bibinfo {title}
  {{Maxwell equations in a curved spacetime: Spin optics approximation}},}\
  }\href {\doibase 10.1103/PhysRevD.102.084013} {\bibfield  {journal} {\bibinfo
   {journal} {Phys. Rev. D}\ }\textbf {\bibinfo {volume} {102}},\ \bibinfo
  {pages} {084013} (\bibinfo {year} {2020})},\ \Eprint
  {http://arxiv.org/abs/2007.03743} {arXiv:2007.03743 [gr-qc]} \BibitemShut
  {NoStop}%
\bibitem [{\citenamefont {Dahal}(2023)}]{Dahal:2022gop}%
  \BibitemOpen
  \bibfield  {author} {\bibinfo {author} {\bibfnamefont {Pravin~Kumar}\
  \bibnamefont {Dahal}},\ }\bibfield  {title} {\enquote {\bibinfo {title}
  {{Covariant formulation of spin optics for electromagnetic waves}},}\ }\href
  {\doibase 10.1007/s00340-022-07952-2} {\bibfield  {journal} {\bibinfo
  {journal} {Appl. Phys. B}\ }\textbf {\bibinfo {volume} {129}},\ \bibinfo
  {pages} {11} (\bibinfo {year} {2023})},\ \Eprint
  {http://arxiv.org/abs/2208.04725} {arXiv:2208.04725 [gr-qc]} \BibitemShut
  {NoStop}%
\bibitem [{\citenamefont {Yamamoto}(2018)}]{Yamamoto:2017gla}%
  \BibitemOpen
  \bibfield  {author} {\bibinfo {author} {\bibfnamefont {Naoki}\ \bibnamefont
  {Yamamoto}},\ }\bibfield  {title} {\enquote {\bibinfo {title} {{Spin Hall
  effect of gravitational waves}},}\ }\href {\doibase
  10.1103/PhysRevD.98.061701} {\bibfield  {journal} {\bibinfo  {journal} {Phys.
  Rev. D}\ }\textbf {\bibinfo {volume} {98}},\ \bibinfo {pages} {061701}
  (\bibinfo {year} {2018})},\ \Eprint {http://arxiv.org/abs/1708.03113}
  {arXiv:1708.03113 [hep-th]} \BibitemShut {NoStop}%
\bibitem [{\citenamefont {Andersson}\ \emph
  {et~al.}(2021{\natexlab{a}})\citenamefont {Andersson}, \citenamefont
  {Joudioux}, \citenamefont {Oancea},\ and\ \citenamefont
  {Raj}}]{Andersson:2020gsj}%
  \BibitemOpen
  \bibfield  {author} {\bibinfo {author} {\bibfnamefont {Lars}\ \bibnamefont
  {Andersson}}, \bibinfo {author} {\bibfnamefont {J\'er\'emie}\ \bibnamefont
  {Joudioux}}, \bibinfo {author} {\bibfnamefont {Marius~A.}\ \bibnamefont
  {Oancea}}, \ and\ \bibinfo {author} {\bibfnamefont {Ayush}\ \bibnamefont
  {Raj}},\ }\bibfield  {title} {\enquote {\bibinfo {title} {{Propagation of
  polarized gravitational waves}},}\ }\href {\doibase
  10.1103/PhysRevD.103.044053} {\bibfield  {journal} {\bibinfo  {journal}
  {Phys. Rev. D}\ }\textbf {\bibinfo {volume} {103}},\ \bibinfo {pages}
  {044053} (\bibinfo {year} {2021}{\natexlab{a}})},\ \Eprint
  {http://arxiv.org/abs/2012.08363} {arXiv:2012.08363 [gr-qc]} \BibitemShut
  {NoStop}%
\bibitem [{\citenamefont {Dahal}(2022)}]{Dahal:2021qel}%
  \BibitemOpen
  \bibfield  {author} {\bibinfo {author} {\bibfnamefont {Pravin~Kumar}\
  \bibnamefont {Dahal}},\ }\bibfield  {title} {\enquote {\bibinfo {title}
  {{Spin Optics for Gravitational Waves}},}\ }\href {\doibase
  10.3390/astronomy1030016} {\bibfield  {journal} {\bibinfo  {journal}
  {Astronomy}\ }\textbf {\bibinfo {volume} {1}},\ \bibinfo {pages} {271--287}
  (\bibinfo {year} {2022})},\ \Eprint {http://arxiv.org/abs/2107.02761}
  {arXiv:2107.02761 [gr-qc]} \BibitemShut {NoStop}%
\bibitem [{\citenamefont {Kubota}\ \emph {et~al.}(2024)\citenamefont {Kubota},
  \citenamefont {Arai},\ and\ \citenamefont {Mukohyama}}]{Kubota:2023dlz}%
  \BibitemOpen
  \bibfield  {author} {\bibinfo {author} {\bibfnamefont {Kei-ichiro}\
  \bibnamefont {Kubota}}, \bibinfo {author} {\bibfnamefont {Shun}\ \bibnamefont
  {Arai}}, \ and\ \bibinfo {author} {\bibfnamefont {Shinji}\ \bibnamefont
  {Mukohyama}},\ }\bibfield  {title} {\enquote {\bibinfo {title} {{Spin optics
  for gravitational waves lensed by a rotating object}},}\ }\href {\doibase
  10.1103/PhysRevD.109.044027} {\bibfield  {journal} {\bibinfo  {journal}
  {Phys. Rev. D}\ }\textbf {\bibinfo {volume} {109}},\ \bibinfo {pages}
  {044027} (\bibinfo {year} {2024})},\ \Eprint
  {http://arxiv.org/abs/2309.11024} {arXiv:2309.11024 [gr-qc]} \BibitemShut
  {NoStop}%
\bibitem [{\citenamefont {Shoom}(2021)}]{Shoom:2020zhr}%
  \BibitemOpen
  \bibfield  {author} {\bibinfo {author} {\bibfnamefont {Andrey~A.}\
  \bibnamefont {Shoom}},\ }\bibfield  {title} {\enquote {\bibinfo {title}
  {{Gravitational Faraday and spin-Hall effects of light}},}\ }\href {\doibase
  10.1103/PhysRevD.104.084007} {\bibfield  {journal} {\bibinfo  {journal}
  {Phys. Rev. D}\ }\textbf {\bibinfo {volume} {104}},\ \bibinfo {pages}
  {084007} (\bibinfo {year} {2021})},\ \Eprint
  {http://arxiv.org/abs/2006.10077} {arXiv:2006.10077 [gr-qc]} \BibitemShut
  {NoStop}%
\bibitem [{\citenamefont {Shoom}(2024)}]{Shoom:2024zep}%
  \BibitemOpen
  \bibfield  {author} {\bibinfo {author} {\bibfnamefont {Andrey~A.}\
  \bibnamefont {Shoom}},\ }\bibfield  {title} {\enquote {\bibinfo {title}
  {{Gravitational Faraday and spin-Hall effects of light: Local
  description}},}\ }\href@noop {} {\  (\bibinfo {year} {2024})},\ \Eprint
  {http://arxiv.org/abs/2404.15934} {arXiv:2404.15934 [gr-qc]} \BibitemShut
  {NoStop}%
\bibitem [{\citenamefont {Frolov}(2024)}]{Frolov:2024ebe}%
  \BibitemOpen
  \bibfield  {author} {\bibinfo {author} {\bibfnamefont {Valeri~P.}\
  \bibnamefont {Frolov}},\ }\bibfield  {title} {\enquote {\bibinfo {title}
  {{Spinoptics in a curved spacetime}},}\ }\href@noop {} {\  (\bibinfo {year}
  {2024})},\ \Eprint {http://arxiv.org/abs/2405.01777} {arXiv:2405.01777
  [gr-qc]} \BibitemShut {NoStop}%
\bibitem [{\citenamefont {Stachel}\ and\ \citenamefont
  {Plebanski}(1977)}]{Stachel:1977cm}%
  \BibitemOpen
  \bibfield  {author} {\bibinfo {author} {\bibfnamefont {J.}~\bibnamefont
  {Stachel}}\ and\ \bibinfo {author} {\bibfnamefont {J.}~\bibnamefont
  {Plebanski}},\ }\bibfield  {title} {\enquote {\bibinfo {title} {{Classical
  Particles with Spin. 1. The WKBJ Approximation}},}\ }\href {\doibase
  10.1063/1.523222} {\bibfield  {journal} {\bibinfo  {journal} {J. Math.
  Phys.}\ }\textbf {\bibinfo {volume} {18}},\ \bibinfo {pages} {2368--2374}
  (\bibinfo {year} {1977})}\BibitemShut {NoStop}%
\bibitem [{\citenamefont {Maccallum}\ and\ \citenamefont
  {Taub}(1972)}]{Maccallum:1972er}%
  \BibitemOpen
  \bibfield  {author} {\bibinfo {author} {\bibfnamefont {Malcolm A.~H.}\
  \bibnamefont {Maccallum}}\ and\ \bibinfo {author} {\bibfnamefont {A.~H.}\
  \bibnamefont {Taub}},\ }\bibfield  {title} {\enquote {\bibinfo {title}
  {{Variational principles and spatially-homogeneous universes, including
  rotation}},}\ }\href {\doibase 10.1007/BF01877589} {\bibfield  {journal}
  {\bibinfo  {journal} {Commun. Math. Phys.}\ }\textbf {\bibinfo {volume}
  {25}},\ \bibinfo {pages} {173--189} (\bibinfo {year} {1972})}\BibitemShut
  {NoStop}%
\bibitem [{\citenamefont {Stein}\ and\ \citenamefont
  {Yunes}(2011)}]{Stein:2010pn}%
  \BibitemOpen
  \bibfield  {author} {\bibinfo {author} {\bibfnamefont {Leo~C.}\ \bibnamefont
  {Stein}}\ and\ \bibinfo {author} {\bibfnamefont {Nicolas}\ \bibnamefont
  {Yunes}},\ }\bibfield  {title} {\enquote {\bibinfo {title} {{Effective
  Gravitational Wave Stress-energy Tensor in Alternative Theories of
  Gravity}},}\ }\href {\doibase 10.1103/PhysRevD.83.064038} {\bibfield
  {journal} {\bibinfo  {journal} {Phys. Rev. D}\ }\textbf {\bibinfo {volume}
  {83}},\ \bibinfo {pages} {064038} (\bibinfo {year} {2011})},\ \Eprint
  {http://arxiv.org/abs/1012.3144} {arXiv:1012.3144 [gr-qc]} \BibitemShut
  {NoStop}%
\bibitem [{\citenamefont {Andersson}\ \emph
  {et~al.}(2021{\natexlab{b}})\citenamefont {Andersson}, \citenamefont
  {Joudioux}, \citenamefont {Oancea},\ and\ \citenamefont {Raj}}]{OOH}%
  \BibitemOpen
  \bibfield  {author} {\bibinfo {author} {\bibfnamefont {Lars}\ \bibnamefont
  {Andersson}}, \bibinfo {author} {\bibfnamefont {J\'er\'emie}\ \bibnamefont
  {Joudioux}}, \bibinfo {author} {\bibfnamefont {Marius~A.}\ \bibnamefont
  {Oancea}}, \ and\ \bibinfo {author} {\bibfnamefont {Ayush}\ \bibnamefont
  {Raj}},\ }\bibfield  {title} {\enquote {\bibinfo {title} {{Propagation of
  polarized gravitational waves}},}\ }\href {\doibase
  10.1103/PhysRevD.103.044053} {\bibfield  {journal} {\bibinfo  {journal}
  {Phys. Rev. D}\ }\textbf {\bibinfo {volume} {103}},\ \bibinfo {pages}
  {044053} (\bibinfo {year} {2021}{\natexlab{b}})},\ \Eprint
  {http://arxiv.org/abs/2012.08363} {arXiv:2012.08363 [gr-qc]} \BibitemShut
  {NoStop}%
\bibitem [{\citenamefont {Barnett}(2014)}]{Barnett:2014era}%
  \BibitemOpen
  \bibfield  {author} {\bibinfo {author} {\bibfnamefont {Stephen~M.}\
  \bibnamefont {Barnett}},\ }\bibfield  {title} {\enquote {\bibinfo {title}
  {{Maxwellian theory of gravitational waves and their mechanical
  properties}},}\ }\href {\doibase 10.1088/1367-2630/16/2/023027} {\bibfield
  {journal} {\bibinfo  {journal} {New J. Phys.}\ }\textbf {\bibinfo {volume}
  {16}},\ \bibinfo {pages} {023027} (\bibinfo {year} {2014})}\BibitemShut
  {NoStop}%
\bibitem [{\citenamefont {Aghapour}\ \emph {et~al.}(2021)\citenamefont
  {Aghapour}, \citenamefont {Andersson},\ and\ \citenamefont
  {Bhattacharyya}}]{Aghapour:2021bkb}%
  \BibitemOpen
  \bibfield  {author} {\bibinfo {author} {\bibfnamefont {Sajad}\ \bibnamefont
  {Aghapour}}, \bibinfo {author} {\bibfnamefont {Lars}\ \bibnamefont
  {Andersson}}, \ and\ \bibinfo {author} {\bibfnamefont {Reebhu}\ \bibnamefont
  {Bhattacharyya}},\ }\bibfield  {title} {\enquote {\bibinfo {title} {{Helicity
  and spin conservation in linearized gravity}},}\ }\href {\doibase
  10.1007/s10714-021-02871-7} {\bibfield  {journal} {\bibinfo  {journal} {Gen.
  Rel. Grav.}\ }\textbf {\bibinfo {volume} {53}},\ \bibinfo {pages} {102}
  (\bibinfo {year} {2021})}\BibitemShut {NoStop}%
\bibitem [{\citenamefont {Mashhoon}\ \emph {et~al.}(2001)\citenamefont
  {Mashhoon}, \citenamefont {Gronwald},\ and\ \citenamefont
  {Lichtenegger}}]{Mashhoon:1999nr}%
  \BibitemOpen
  \bibfield  {author} {\bibinfo {author} {\bibfnamefont {Bahram}\ \bibnamefont
  {Mashhoon}}, \bibinfo {author} {\bibfnamefont {Frank}\ \bibnamefont
  {Gronwald}}, \ and\ \bibinfo {author} {\bibfnamefont {Herbert I.~M.}\
  \bibnamefont {Lichtenegger}},\ }\bibfield  {title} {\enquote {\bibinfo
  {title} {{Gravitomagnetism and the clock effect}},}\ }\href {\doibase
  10.1007/3-540-40988-2_5} {\bibfield  {journal} {\bibinfo  {journal} {Lect.
  Notes Phys.}\ }\textbf {\bibinfo {volume} {562}},\ \bibinfo {pages} {83--108}
  (\bibinfo {year} {2001})},\ \Eprint {http://arxiv.org/abs/gr-qc/9912027}
  {arXiv:gr-qc/9912027} \BibitemShut {NoStop}%
\bibitem [{\citenamefont {Chandrasekhar}(1983)}]{Chandra}%
  \BibitemOpen
  \bibfield  {author} {\bibinfo {author} {\bibfnamefont {S.}~\bibnamefont
  {Chandrasekhar}},\ }\href@noop {} {\emph {\bibinfo {title} {The Mathematical
  Theory of Black Holes}}}\ (\bibinfo  {publisher} {Oxford University Press},\
  \bibinfo {year} {1983})\BibitemShut {NoStop}%
\end{thebibliography}

%\begin{thebibliography}{99}

%\end{thebibliography}


\end{document}